# The AI&M Procedure for Learning from Incomplete Data

**Manfred Jaeger**
Institut for Datalogi, Aalborg Universitet, Fredrik Bajers Vej 7E, DK-9220 Aalborg Ø
jaeger@cs.aau.dk

## Abstract

We investigate methods for parameter learning from incomplete data that is not missing at random. Likelihood-based methods then require the optimization of a profile likelihood that takes all possible missingness mechanisms into account. Optimizing this profile likelihood poses two main difficulties: multiple (local) maxima, and its very high-dimensional parameter space. In this paper a new method is presented for optimizing the profile likelihood that addresses the second difficulty: in the proposed AI&M (adjusting imputation and maximization) procedure the optimization is performed by operations in the space of data completions, rather than directly in the parameter space of the profile likelihood. We apply the AI&M method to learning parameters for Bayesian networks. The method is compared against conservative inference, which takes into account each possible data completion, and against EM. The results indicate that likelihood-based inference is still feasible in the case of unknown missingness mechanisms, and that conservative inference is unnecessarily weak. On the other hand, our results also provide evidence that the EM algorithm is still quite effective when the data is not missing at random.

## 1 INTRODUCTION

Most commonly used methods for learning from incomplete data are based on the assumption that values are missing at random (*mar* [13]). The concept of multivariate data with missing values has been extended to the more general notion of coarse data [4]. The missing at random assumption here has a counterpart called coarsened at random (*car*).

Under the *mar/car* assumptions one can ignore the mechanism that causes complete data cases to become incompletely reported, and statistical inference can be based on the *face-value likelihood* [2] that measures the probability of an incomplete observation by its marginal probability according to the underlying complete data distribution. Methods like the EM algorithm are usually quite effective for maximizing the face-value likelihood.

Very often, however, *mar/car* appear to be rather unsafe assumptions for the data coarsening process, and the question arises what statistical inferences are possible without making these assumptions. One can find in the literature two main approaches to this problem. The first approach is to explicitly model the coarsening mechanism by a parametric model for the distribution of incomplete observations given complete underlying data cases. Such parametric models can represent more or less restrictive assumptions on the coarsening mechanism, including completely unrestricted models. This kind of approach has mostly been suggested for low-dimensional problems, e.g. the analysis of relatively small contingency tables, where the parametric model for the coarsening process then consists of only a small number of additional parameters [7, 10].

The second approach is based on the view that when no assumptions on the coarsening mechanism shall be made, then one should consider all parameter estimates obtained from any possible completion of the data. This will lead to a set estimate for the parameters (typically given in terms of an interval estimate for each parameter), which then may be further refined to a point estimate [8, 9, 11, 1, 15]. This approach, which we call the conservative approach, has been suggested, in particular, for learning parameters in Bayesian networks [12, 1]. In the conservative approach no attempt is made to score the likelihood of different data completions. It is one of the objectives of this paper to argue that this leads to unnecessarily weak inferences, because even when a-priori no assumptions on the data coarsening mechanism are made, the data can still determine that some mechanisms are more likely than others, which then leads to more specific estimates for the complete data parameters.

Our approach, thus, follows the first mentioned line of work, and our goal is to extend this approach to high-dimensional models, without requiring the specification of a restricted parametric model for the coarsening mechanism.

While pursuing this goal, one has to be aware of two fundamental limitations that one cannot hope to overcome: first, in most cases the parameters of interest will not be identifiable from incomplete data, i.e. even in the limit of infinite data the likelihood functions that we encounter have multiple global maxima. This problem already exists for inference with the face-value likelihood under the *mar/car* assumption, but it is exacerbated when no assumption on the coarsening mechanism are made. Second, when no assumptions on the complete data distribution are made, then the data can never refute the *mar/car* assumption [3], i.e. based on the data one cannot infer whether the *mar/car* assumption is reasonable, and thus cannot decide whether methods based on *mar/car* will be appropriate, or other methods should be applied. This result is summarized as "*car* is everything" in [3]. However, as pointed out in [5], this is no longer true when for the complete data distribution a restricted parametric model is assumed. For the high-dimensional data we are concerned with, such a parametric model will always be required.

## 2 COARSE DATA

Throughout this paper we will use the general coarse data model. This is not so much for the sake of added generality, than for the sake of greater simplicity, both conceptually and notationally. In this section we introduce the basic concepts of coarse data, parametric models for the joint distribution of complete and coarse data, and the two central likelihood functions we need to consider.

The underlying complete data is represented by a random variable $X$ with values in a finite state space $W = \{x_1, \ldots, x_n\}$. $X$ typically will be a multi-variate random variable, in which case $W$ is the Cartesian product of the state spaces of the components of $X$. $X$ has a distribution $P_\theta$ for some $\theta$ in a parameter space $\Theta$.

The value of $X$ is observed only incompletely. In the general coarse data model such incomplete observations of $X$ can be given by any subset of the state space $W$. Formally, these observations are the values of a random variable $Y$ with state space $2^W$. We denote $2^W$ with $\mathcal{Y}$ when we want to distinguish it as the sample space of $Y$. It is assumed that the observations $Y$ always contain the true value of $X$ (i.e. the data is incomplete, not incorrect). The joint distribution of $X$ and $Y$, then can be parameterized by $P_\theta$ and parameters

$$\lambda_{x,U} = P(Y = U \mid X = x) \quad (x \in W, U \in \mathcal{Y} : x \in U).$$

Thus, the parameter space of all possible coarsening mechanisms (for the given state space $W$) is

$$\Lambda_{sat} := \{(\lambda_{x,U})_{x \in W, U \in \mathcal{Y} : x \in U} \mid \forall x \in W : \sum_{U : x \in U} \lambda_{x,U} = 1\}.$$

The joint distribution for $(X, Y)$ given by $\theta \in \Theta$ and $\lambda \in \Lambda_{sat}$ is denoted $P_{\theta,\lambda}$. The parameter space $\Lambda_{sat}$ represents the *saturated* coarsening model, i.e. the one that does not encode any assumptions on how the data is coarsened.

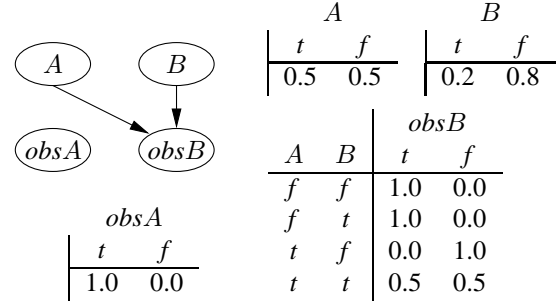

Figure 1: Basic Example

**Example 2.1** *Figure 1 shows a Bayesian network with two binary nodes $A, B$, and two binary observation nodes obsA,obsB. The distribution of interest here is the joint distribution of $A$ and $B$, i.e. in our general terminology: $X = (A, B)$ and $W = \{t, f\} \times \{t, f\}$. The distribution of $X$ is parameterized by $\Theta = \{(\theta_A, \theta_B) \mid \theta_A, \theta_B \in [0, 1]\}$, where e.g. $\theta_A := P(A = t)$. The observation nodes represent a coarsening mechanism, which, in this case, is of the special type of a missing value mechanism. E.g. obsA$= t$ means that the value of $A$ is observed. According to the model, $A$ always is observed, and $B$ can only be unobserved when $A = t$. The model only allows for four distinct observations. The observations, their representation as subsets $U \subseteq W$, and their probabilities are:*

| Observation | $U$ | $P(Y = U)$ |
|---|---|---|
| $A = t, B = ?$ | $U_1 = \{(t, f), (t, t)\}$ | 0.45 |
| $A = t, B = t$ | $U_2 = \{(t, t)\}$ | 0.05 |
| $A = f, B = t$ | $U_3 = \{(f, t)\}$ | 0.1 |
| $A = f, B = f$ | $U_4 = \{(f, f)\}$ | 0.4 |

*For $x \in W$ and $U = U_1, \ldots, U_4$ the $\lambda_{x,U}$ parameters are given by:*

| $x$ | $U_1$ | $U_2$ | $U_3$ | $U_4$ |
|---|---|---|---|---|
| $\{f, f\}$ | nd | nd | nd | 1 |
| $\{f, t\}$ | nd | nd | 1 | nd |
| $\{t, f\}$ | 1 | nd | nd | nd |
| $\{t, t\}$ | 0.5 | 0.5 | nd | nd |

*Entries "nd" mean that the parameter is undefined. All other $\lambda_{x,U}$ parameters are zero.*

Specific assumptions on the coarsening mechanism can be made by delimiting admissible $\lambda$-parameters to some sub-

set of $\Lambda_{sat}$. The *car* assumption corresponds to the subset

$$\Lambda_{car} := \{\lambda \in \Lambda_{sat} \mid \forall U \forall x, x' \in U : \lambda_{x,U} = \lambda_{x',U}\}.$$

The coarsening mechanism of Example 2.1 is not *car*, because for $U = U_1$, $x = \{t, f\}$, $x' = \{t, t\}$ we have $1 = \lambda_{x,U} \neq \lambda_{x',U} = 0.5$.

A particular coarsening model $\Lambda \subseteq \Lambda_{sat}$ induces a likelihood function on the parameter space $\Theta$ by maximizing over $\lambda$-values. Thus, given a sample $\boldsymbol{U} = (U_1, \ldots, U_N)$ of $Y$, one obtains the *profile($\Lambda$)-likelihood*

$$L_\Lambda(\theta \mid \boldsymbol{U}) := \max_{\lambda \in \Lambda} \prod_{i=1}^{N} P_{\theta,\lambda}(Y = U_i).$$

Alternatively, in a fully Bayesian analysis, one can also consider the likelihood on $\Theta$ obtained by integrating, rather than maximizing, over $\Lambda$. In this paper we will only be concerned with $\Lambda = \Lambda_{sat}$ (no assumptions on the coarsening mechanism) and $\Lambda = \Lambda_{car}$ (*car* assumption). We call the resulting profile likelihoods simply *profile(sat)-*, respectively *profile(car)-likelihood*, denoted $L_{sat}, L_{car}$. We also write $LL_{sat}$ and $LL_{car}$ for the corresponding log-likelihoods.

The result that under the *car* assumption the coarsening mechanism can be ignored derives from the fact that the profile(car)-likelihood factors as

$$L_{car}(\theta \mid \boldsymbol{U}) = f(\boldsymbol{U}) L_{FV}(\theta \mid \boldsymbol{U}),$$

where

$$f(\boldsymbol{U}) := \max_{\lambda \in \Lambda_{car}} \prod_{i=1}^{N} \lambda_{U_i},$$

and $L_{FV}$ is the face-value likelihood [2]

$$L_{FV}(\theta \mid \boldsymbol{U}) := \prod_{i=1}^{N} P_\theta(X \in U_i).$$

This is essentially Rubin's original result [13]. Since $f(\boldsymbol{U})$ does not depend on $\theta$, it establishes that under the *car* assumption inference can be based on the face-value likelihood (some subtle difficulties arise from the fact that $P_\theta$ may have a varying sets of support for different $\theta \in \Theta$; we ignore these issues in this paper, and refer to [5] for a detailed discussion). We conclude this section by re-stating in our framework what is essentially the "car is everything" result of [3].

**Theorem 2.2** *Let $\Theta_{\text{sat}}$ be a parameter space that contains for every possible distribution $P$ on $W$ a parameter $\theta$ with $P_\theta = P$. Let $\boldsymbol{U}$ be incomplete data for $X$, and $\hat{\theta} \in \Theta_{\text{sat}}$ a global maximum of $L_{\text{car}}(\cdot \mid \boldsymbol{U})$. Then $\hat{\theta}$ also is a global maximum of $L_{\text{sat}}(\cdot \mid \boldsymbol{U})$.*

An important implication of this theorem is that under the model $\Theta_{sat}$ *car* cannot be tested. The theorem does not hold any longer when for $X$ a restricted parameterization $\Theta$ is assumed [5]. Thus, we can say informally: '*car*' cannot be tested against '*not car*', but '*car* and $\theta \in \Theta$' can (sometimes) be tested against '*not car* and $\theta \in \Theta$'.

## 3 OPTIMIZING $L_{sat}$

Any profile($\Lambda$)-likelihood function can, in principle, be used for any kind of statistical inference relying on a likelihood function, including Bayesian updating of a parameter prior. However, in this paper we shall only be concerned with maximum-likelihood inference, i.e. finding a parameter $\hat{\theta}$ maximizing the profile($\Lambda$)-likelihood. Under the *car* assumption this reduces to maximizing the face-value likelihood, for which methods like expectation maximization (EM) or multiple imputation (MI) are quite effective methods.

It is known that any profile($\Lambda$)-likelihood can be optimized by interpreting observations of $Y$ as incomplete observations of $Z := (X, Y)$. Interpreted in this way, the data is *car*, and one can employ e.g. the EM algorithm to find a maximum likelihood parameter $(\hat{\theta}, \hat{\lambda})$ for $Z$. The $\hat{\theta}$ component of this solution then is a maximum of the profile($\Lambda$)-likelihood [6]. However, this method requires an optimization over the parameter space $\Theta \times \Lambda$. For $\Lambda = \Lambda_{sat}$ this becomes infeasible very quickly, as the number of parameters in $\Lambda_{sat}$ is exponential in the size of the state space $W$. In this section, therefore, we propose a new general method for optimizing $L_{sat}$, which avoids an explicit optimization over $\Lambda_{sat}$. Before introducing the method, we illustrate by our basic example that optimizing $L_{sat}$ can actually lead to the desired results.

**Example 3.1** *(Example 2.1 continued) Suppose we have a large representative sample $\boldsymbol{U}$ for $Y$, i.e. the empirical distribution $\hat{P}$ defined by the sample has the expected values: $\hat{P}(U_1) = 0.45, \ldots, \hat{P}(U_4) = 0.4$. One can show by elementary means that the profile(sat)-likelihood then has a unique maximum at the correct parameter $\theta_0 = (0.5, 0.2)$. Basically, this is due to the fact that $\theta_0$ is the only parameter in $\Theta$ for which there exists $\lambda_0 \in \Lambda_{sat}$, such that the resulting distribution $P_{\theta_0 \lambda_0}$ has the observed empirical marginal on $\mathcal{Y}$.*

*Conservative inference will in this example lead to the bounds $\theta_A = 0.5, \theta_B \in [0.15, 0.6]$. The refinement operations proposed in [12, 1] would furthermore select the center point 0.375 of [0.15,0.6] as the point estimate for $\theta_B$.*

*We can also compute analytically the (unique) maximum of the face-value likelihood, which turns out to be $\theta_1 = (0.5, 0.2727)$. Finally, in this simple example we can com-*

pute the values of the two profile likelihoods

$$LL_{\text{sat}}(\theta_0 \mid \boldsymbol{U}) = -1.1059, \ LL_{\text{car}}(\theta_1 \mid \boldsymbol{U}) = -1.1779.$$

*Thus, a likelihood ratio test would indicate that the observed data is not missing at random.*

In the preceding example all necessary computations could be carried out explicitly in the parameter space $\Theta \times \Lambda_{sat}$. As noted above, this becomes quickly infeasible as the size of $W$ increases. As the main contribution in this paper, we are now going to develop a general approach for optimizing $LL_{sat}$ that works in the space of data completions, rather than directly in $\Lambda_{sat}$. The following definition introduces our concept of data completion, which allows for fractional completions, i.e. the probability mass of one incomplete observation can be distributed over several of its possible completions.

**Definition 3.2** *Let $\boldsymbol{U} = U_1, \ldots, U_N$ be a dataset. A completion of $\boldsymbol{U}$ is a mapping $c$ that assigns to every $U_i \in \boldsymbol{U}$ a probability distribution $c(U_i)$ over $U_i$. The completion $c$ defines a probability distribution $P_c := 1/N \sum_{i=1}^{N} c(U_i)$ on $W$. When $c(U_i)(x) = 1$ for some $x \in W$, then $c$ is called a 1-completion, and we also write shortly $c(U_i) = x$. We denote with $\mathcal{C}(\boldsymbol{U})$ the set of all completions of $\boldsymbol{U}$.*

The following theorem provides an alternative representation of the profile(sat)-likelihood in terms of data completions. $H(\cdot)$ here denotes entropy, and $CE(\cdot, \cdot)$ is cross-entropy distance.

**Theorem 3.3** *Let $\boldsymbol{U} = U_1, \ldots, U_N$ be a dataset, and $m$ the empirical distribution defined by $\boldsymbol{U}$ on $\mathcal{Y}$.*

$$\frac{1}{N} LL_{\text{sat}}(\theta \mid \boldsymbol{U}) = H(m) - \min_{c \in \mathcal{C}(\boldsymbol{U})} CE(P_c, P_\theta).$$

The proof of the theorem is quite straightforward, based on well-known structural properties of cross-entropy.

According to Theorem 3.3, an optimal parameter $\hat{\theta}$ for $LL_{sat}$ is equivalently characterized as

$$\hat{\theta} = \underset{\theta \in \Theta}{arg \, min} \min_{c \in \mathcal{C}(\boldsymbol{U})} CE(P_c, P_\theta).$$

Based on this characterization, we propose a general procedure for optimizing $LL_{sat}$. The procedure is given in Table 1.

The AI&M procedure bears much resemblance to the EM procedure. In the AI step (for adjusting imputation) a data completion is computed, such that the distribution defined by the completion fits the current estimate $P_{\theta_t}$ as closely as possible. The E step of the EM procedure, on the other hand, corresponds to setting $c_t(\boldsymbol{U})$ to the expected completion given the current $P_{\theta_t}$. The M step, then, is the same in both procedures: given the current data completion, $\theta_{t+1}$

$t := 0$
Choose initial $\theta_0 \in \Theta$
**repeat**
    $c_t := arg \, min_{c \in \mathcal{C}(\boldsymbol{U})} CE(P_c, P_{\theta_t})$     (AI step)
    $\theta_{t+1} := arg \, min_{\theta \in \Theta} CE(P_{c_t}, P_\theta)$     (M step)
    $t := t + 1$
**until** <*termination condition*>.

Table 1: The AI&M procedure

is set to the maximum likelihood parameter given the complete data. In the AI&M procedure, this maximization of likelihood is expressed by an equivalent minimization of cross-entropy. The AI&M iterations are continued until some termination condition applies. A suitable condition is that $CE(P_{c_{t-1}}, P_{\theta_t}) - CE(P_{c_t}, P_{\theta_{t+1}})$ is smaller than some threshold. It is straightforward to verify that AI&M iterations are score improving for the profile(sat)-likelihood:

$$LL_{sat}(\theta_{t+1} \mid \boldsymbol{U}) \geq LL_{sat}(\theta_t \mid \boldsymbol{U}). \quad (1)$$

This alone does not guarantee that the sequence $\theta_t$ converges to a local maximum (or even a saddle-point) of $LL_{sat}$. The only immediate conclusion we can draw is that the sequence of $LL_{sat}(\theta_t \mid \boldsymbol{U})$ values will converge, and that a termination condition like the one mentioned above will eventually apply. An investigation of the exact convergence properties of the the AI&M procedure is outside the scope of this paper. However, it is conjectured that similar regularity conditions as needed to ensure convergence of the EM algorithm [14] will also guarantee convergence of AI&M to a stationary point of $LL_{sat}$.

Rephrasing the maximization of $L_{sat}$ in terms of an optimization over data completions does not immediately mean a simplification of the problem, since the space $\mathcal{C}(\boldsymbol{U})$ is just as intractable as the space $\Lambda_{sat}$. However, the reduction of the problem to performing the AI step in the AI&M procedure (the M step typically being easy), provides opportunities for developing for specific types of models exact or approximate efficient implementations of the AI step. In the following section we shall consider a quite simple approximate implementation of the AI step for parameter learning for Bayesian networks.

## 4 BAYESIAN NETWORKS

### 4.1 APPROXIMATE AI FOR BAYESIAN NETWORKS

Our approximate solution for the AI step is based on two main elements: first, we work only with 1-completions of the data; second, we conduct the search for optimal 1-completions in an iterative process in which a current candidate completion is modified by changing for one incomplete data case the value of one unobserved compo-nent. In order to mitigate the limitations imposed by the restriction to 1-completions, we initially replace the dataset $U_1, \ldots, U_N$ with a dataset $U_1, \ldots, U_{zN}$, where each original data case is replaced by $z$ copies of itself. A 1-completion of this new dataset then corresponds to a fractional completion of the old dataset, where the probability mass of an incomplete $U_i$ can be distributed in the form $c(U_i)(x) = l/z$ ($l \in \mathbb{N}, 1 \leq l \leq z$) to at most $z$ different states $x \in U_i$. Higher values of $z$ mean a more accurate implementation of the AI step, at a higher computational cost.

1  **Input:** Incomplete observations $U_1, \ldots, U_{zN}$ of variables $V_1, \ldots, V_k$ ($z$ copies of $N$ data cases). Previous completion $c_{t-1}$; current parameters $\theta_t$.
2  $c_{t,0} := c_{t-1}$
3  **for** $j = 1, \ldots, zN$
4   Let $x^{(0)} := c_{t,j-1}(U_j)$
5   Let $x^{(1)}, \ldots, x^{(l)}$ be the set of 1-completions of $U_j$ that differ from $x^{(0)}$ in exactly one component
6   Let $c^{(1)}, \ldots, c^{(l)}$ be the 1-completions obtained from $c_{t,j-1}$ by replacing $x^{(0)}$ with $x^{(1)}, \ldots, x^{(l)}$.
7   Set $c_{t,j} := \arg\min_{c \in \{c_{t,j-1}, c^{(1)}, \ldots, c^{(l)}\}} CE(P_c, P_{\theta_t})$
8  **return** $c_t := c_{t,zN}$

Table 2: Approximate AI step for Bayesian networks

Table 2 describes the approximate AI step in greater detail. The maximization in step 7 of the algorithm is a "local" computation: the value $CE(P_{c_{t,j-1}}, P_{\theta_t})$ is given at this point. The new candidate completions $c^{(i)}$ for $c_{t,j}$ differ from $c_{t,j-1}$ for only the two states $x^{(0)}$ and $x^{(i)}$, and the new CE-values can be computed by only re-computing the contributions of these two states to the CE-values. The algorithm guarantees that $CE(P_{c_t}, P_{\theta_t}) \leq CE(P_{c_{t-1}}, P_{\theta_t})$, and therefore (1) still holds under this approximate AI step.

The AI&M procedure with the approximate AI step has been implemented in Java, using the Hugin (www.hugin.com) system for all standard Bayesian network computations, including the EM computations referred to in section 4.4.

### 4.2  EXPERIMENTAL SETUP

| Name | # nodes | # edges | $|W|$ |
|---|---|---|---|
| Basic | 2 | 0 | 4 |
| Asia | 8 | 8 | 256 |
| Alarm/Alarm(R) | 37 | 46 | $1.7 \cdot 10^{16}$ |

Table 3: Bayesian networks used

All our experiments are based on given Bayesian network models for the underlying complete data. Four different models were used. Table 3 summarizes some of their features. Basic is the 2-node network consisting of the primary nodes $A, B$ in Figure 1. Asia and Alarm are the standard Bayesian network models of theses names, as distributed with the Hugin package. Alarm(R) is the Alarm network with all parameters in the conditional probability tables replaced by uniformly sampled random entries.

The experiments are based on samples of incomplete data generated as follows:

1 Choose parameters $mp, N \in \mathbb{N}, \mu, \sigma \in [0,1]$.

2 Let $B$ be a Bayesian network with nodes $V_1, \ldots, V_k$ and parameters $\theta_B$.

3 Add to $B$ binary nodes $obsV_i$ ($i = 1, \ldots, k$).

4 For each $i$ select randomly an integer $k_i$ between 0 and $mp$. Let $Pa_i$ be a set of $k_i$ nodes selected randomly and uniformly from $V_1, \ldots, V_k, obsV_1, \ldots, obsV_{i-1}$. Connect $Pa_i \cup \{V_i\}$ as parents to $obsV_i$.

5 For each parent configuration *conf* of $obsV_i$ set the conditional probability for $P(obsV_i = false \mid conf)$ to a value randomly sampled from the Beta distribution with mean $\mu$ and variance $\sigma$.

6 Sample $N$ instantiations in the extended Bayesian network. Turn this sample into an incomplete sample for $V_1, \ldots, V_k$ by deleting the value for $V_i$ iff $obsV_i = false$.

With the $\mu$ and $\sigma$ parameters we can control two relevant aspects of the resulting coarsening model: $\mu$ is the expected percentage of missing values in large samples from the random coarsening mechanisms; $\sigma$ controls, to some extent, how far from *mar* the generated data is: large values of $\sigma$ mean that the cpt-rows of observation nodes contain mostly entries close to 0 and 1, which means a strong dependence on the parents, and thus a strong possibility for non-*mar* patterns. Setting $\sigma = 0$, on the other hand, means that the dependence on the observation nodes on the parents is spurious, and the data is actually *mar*. In our experiments we mostly use $\mu = 0.1$ and $\sigma = 0.05$. It has turned out that for the smaller networks this setting already leads to quite a diversity in the obtained incomplete data distributions, with the percentage of missing values ranging from about 2% to 25%.

Given the network structure of $B$ and a sample $U_1, \ldots, U_N$ generated by this procedure, it will be our goal to estimate the parameters $\theta_B$. We use two different measures for the quality of an estimate $\hat{\theta}$ for $\theta_B$. First, we consider $CE(P_{\theta_B}, P_{\hat{\theta}})$. This is the statistically most pertinent measure, as it corresponds to the likelihood of $\hat{\theta}$ given a large dataset of complete cases sampled from $P_{\theta_B}$. However, $CE(P_{\theta_B}, P_{\hat{\theta}})$ can be somewhat problematic, because it will be infinite when $\hat{\theta}$ contains zero-components that are non-zero in $\theta_B$. In order to avoid this problem, we will always apply a parameter smoothing operation to an initially obtained $\hat\theta$ (specifically, we smooth the components of $\hat\theta$ by replacing $\hat\theta_i$ with $(\hat\theta_i k+1)/(k+m)$, where $k$ is the number of data cases from which the cpt-row of $\hat\theta_i$ was estimated, and $m$ is the number of components in that row; this corresponds to adding a pseudo-count of one to each cell in each cpt). The same smoothing operation is applied to the results of all learning procedures we consider. In order to obtain a comparison that is not so strongly affected by how accurately small parameters are estimated, we also evaluate $\hat\theta$ by its mean squared error relative to $\theta_B$.

### 4.3 AI&M VS. CONSERVATIVE LEARNING

Our first experiments are aimed at finding whether optimizing $L_{sat}$ via AI&M leads to substantially different (and better) results than conservative inference. This would not be the case, for instance, if $L_{sat}$ had so many global maxima that any parameter $\hat\theta$ obtained from some data completion lies close to a maximum of $L_{sat}$.

For this experiment the Asia and Alarm networks were used. Eight (Asia), respectively five (Alarm) incomplete datasets of size 1000 were generated according to our general procedure ($\mu = 0.1, \sigma = 0.05$). For each dataset, 10 different random completions were generated, and the (smoothed) estimates $\theta_1^{cons},\ldots,\theta_{10}^{cons}$ were computed. Under conservative inference, each $\theta_i^{cons}$ is an element of the set estimate $\hat\Theta$ for $\theta_B$. The $\theta_i^{cons}$ were then used as initial points for the AI&M procedure, which was run until convergence to some point $\theta_i^{aim}$. For $i=1,\ldots,10$ then $CE(P_{\theta_B},P_{\theta_i^{cons}})$ and $CE(P_{\theta_B},P_{\theta_i^{aim}})$ were computed.

A reduction of the *CE*-values when going from $\theta^{cons}$ to $\theta^{aim}$ indicates that the data allows identification of the true parameter $\theta$ beyond the conservative estimate $\hat\Theta$, and that the AI&M procedure is successful in finding more accurate estimates. Figure 2 shows the results obtained by plotting the $CE(P_{\theta_B},P_{\theta_i^{cons}})$-values on the $x$-axis against the $CE(P_{\theta_B},P_{\theta_i^{aim}})$-values. The different datasets here are labeled by the percentage of missing values they contain. The results show that for all five datasets from the Alarm network, and for at least three datasets from the Asia network (labeled 7%,12.5%,6.8%) the AI&M procedure consistently leads to a substantially better estimate than the original starting point. For 4 out of the remaining 5 Asia datasets we still observe an improvement, but the quality of the final estimate $\theta_i^{aim}$ varies more widely. For one dataset (3.3%) the AI&M procedure actually leads to worse estimates. This could happen, for example, when most local maxima of $L_{sat}$ happen to lie far away from the true $\theta_B$.

### 4.4 AI&M VS. EM

Even though the data we are generating is not *mar*, one also needs to compare the results obtained by AI&M with results obtained by EM. The question here basically is whether EM, while based on wrong assumptions, may not be more successful in identifying the true parameter, because the profile(car)-likelihood it optimizes has fewer local maxima than the profile(sat)-likelihood, and because its implementation provides a more exact optimization procedure than the approximate implementation of AI&M. The results we will observe in these experiments clearly can depend very much on our chosen method for generating non-*mar* data. If the datasets we produce for some reason tend to be "almost *mar*", then EM can be expected to produce better results than AI&M. While our data generating procedure has been designed so as to produce data that clearly is not *mar*, it is also certainly the case that we are exploring only a relatively small part of the universe of possible coarsening mechanisms, and that experimental results might change significantly for other data generating strategies.

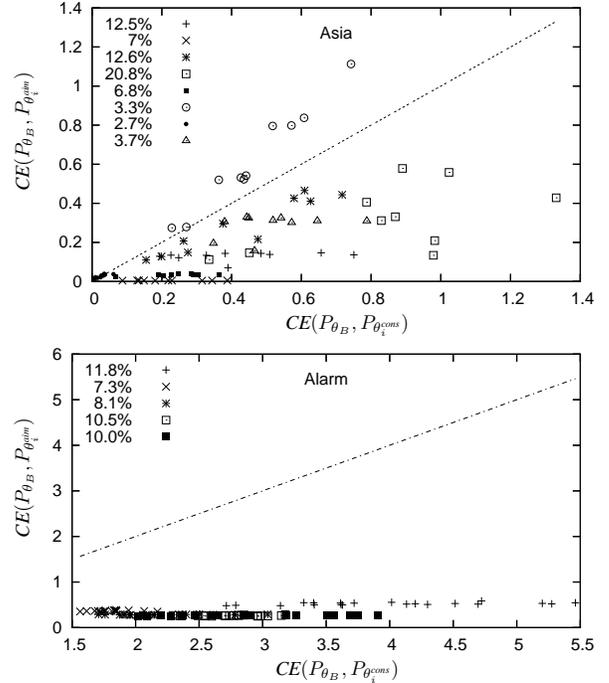

Figure 2: AI&M vs. Conservative Learning: Asia and Alarm

All our following experiments are conducted as follows: under the experiment specific setting of parameters 100 incomplete datasets (from 100 different coarsening models) are generated. For each dataset, first the EM estimate $\theta_i^{em}$ ($i=1,\ldots,100$) is computed. The $\theta_i^{em}$ are then used as initial points for the AI&M procedure, which is run until convergence to some point $\theta_i^{aim}$. The results are evaluated using the differences $CE(P_{\theta_B},P_{\theta_i^{aim}}) - CE(P_{\theta_B},P_{\theta_i^{em}})$ and $MSE(P_{\theta_B},P_{\theta_i^{aim}}) - MSE(P_{\theta_B},P_{\theta_i^{em}})$. Thus, negative numbers mean that AI&M found a better estimate of the true parameter; positive numbers show an advantage of EM.

| Modifications | Model | Coarsening | $N$ | $z$ | CE-final | CE-diff. | MSE-diff | Score |
|---|---|---|---|---|---|---|---|---|
| Base | Asia | 2:0.1:0.05 | 1000 | 5 | 0.096±0.190 | -0.029±0.084 | -0.001±0.006 | 0.011± 0.005 |
| Sample Size | Asia | 2:0.1:0.05 | 500 | 5 | 0.149±0.222 | -0.016±0.091 | 0.0±0.013 | 0.018± 0.008 |
| · | Asia | 2:0.1:0.05 | 2000 | 5 | 0.141±0.247 | -0.053±0.154 | -0.002±0.006 | 0.006± 0.003 |
| · | Asia | 2:0.1:0.05 | 5000 | 5 | 0.119±0.278 | -0.034±0.210 | -0.002±0.005 | 0.003± 0.001 |
| *mp* parameter | Asia | 0:0.1:0.05 | 1000 | 5 | 0.218±0.356 | 0.005±0.076 | 0±0.006 | 0.01± 0.005 |
| · | Asia | 8:0.1:0.05 | 1000 | 5 | 0.071±0.117 | -0.034±0.083 | -0.001±0.005 | 0.011± 0.005 |
| $z$ parameter | Asia | 2:0.1:0.05 | 1000 | 1 | 0.145±0.300 | -0.007±0.118 | 0.001±0.005 | 0.014± 0.005 |
| · | Asia | 2:0.1:0.05 | 1000 | 10 | 0.162±0.267 | -0.040±0.142 | 0.0±0.007 | 0.01± 0.005 |
| $\sigma$ parameter | Asia | 2:0.1:0.01 | 1000 | 5 | 0.033±0.023 | 0.014±0.018 | -0.001±0.003 | 0.007± 0.003 |
| · | Asia | 2:0.1:0.00 | 1000 | 5 | 0.028±0.012 | 0.017±0.012 | -0.001±0.005 | 0.006± 0.002 |
| Models | Basic | Fig. 1 | 1000 | 5 | 0.002±0.002 | -0.016±0.008 | -0.003±0.002 | 0±0 |
| · | Alarm | 5:0.2:0.05 | 1000 | 10 | 0.334±0.051 | -0.010±0.016 | 0.001±0.001 | 2.227±0.130 |
| · | Alarm(R) | 5:0.1:0.05 | 1000 | 10 | 1.016±0.569 | 0.168±0.197 | 0.001±0.001 | 21.497±0.393 |

Table 4: Results: AI&M vs. EM

We start with an experiment using data sampled from the Asia network with settings $mp = 2$, $\mu = 0.1, \sigma = 0.05$ and the AI&M procedure run with $z = 5$. Figure 3 shows the results. The pattern we here observe is quite characteristic for most of the experiments: in most cases EM and AI&M produce estimates of very similar quality (indicating that AI&M did not venture very far from the EM starting point). Nevertheless, there appears to be a small but clear tendency for AI&M to produce better results. The overall evaluation is made more difficult by the existence of a few rather extreme outliers (which can be both in favor of EM or AI&M), which lead to a high variance in the results.

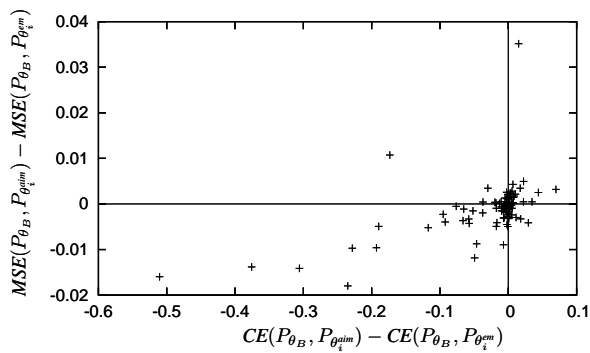

Figure 3: Asia base experiment

Table 4 gives additional details on this experiment (in row labeled 'Base'), as well as several other experiments in which some experimental parameters were modified. The column 'Coarsening' gives the parameters of the incomplete data generator in the format $mp : \mu : \sigma$. $N$ is the sample size, and $z$ the data-duplication parameter of the approximate AI step. The columns 'CE-diff' and 'MSE-diff' give the mean and standard deviation (always for 100 runs of the experiment) of our already described evaluation measure. 'CE-final' gives mean and standard deviation for $CE(P_{\theta_B}, P_{\theta_i^{aim}})$, and 'Score' is $CE(c_t, P_{\theta_t})$ at termination of the AI&M procedure. Note that by Theorem 3.3 a zero score here means that $\theta_t$ is a global maximum of $L_{sat}$. This also implies that when the score values are close to zero, no significantly different results can be expected from an exact implementation of the AI step, since our approximate implementation is already successful in finding an estimate $\theta^{aim}$ close to a global maximum of $L_{sat}$.

The results show that AI&M maintains in most experiments with the small networks its slight advantage over EM. EM gains an advantage when the $\sigma$ parameter is reduced. However, even for *mar*-data ($\sigma = 0$) the difference is not very large. The case $mp = 0$ is not very easy to analyze. On the one hand, the data here is extremely non-*mar* (missingness for variable $V_i$ only depends on the value of $V_i$). On the other hand, this mechanism should also be particularly hard to identify by AI&M. It is not surprising, therefore, that both EM and AI&M showed the worst performance on this version of the Asia experiments.

The experiment with the Basic network differed from the other experiments in that for the generation of all 100 datasets the same coarsening model as described in Example 2.1 was used. The results here show that the AI&M procedure provides quite consistently accurate estimates of the true parameters. Furthermore, the CE-diff result is close to the theoretical value $CE(P_{\theta_B}, P_{\theta_0}) - CE(P_{\theta_B}, P_{\theta_1}) = -0.014$ (with $\theta_0, \theta_1$ as in Example 3.1).

The experiment for the Alarm network shows that AI&M also works on large state spaces. Both EM and AI&M gave poorer results on Alarm(R), with EM gaining a considerable advantage. The problem for AI&M here seems to be that the true distribution is relatively evenly spread over the state space (whereas Alarm contains many near-zero entries, leading to a somewhat concentrated distribution). The approximate AI step, under the settings in this experiment, produces data completions that define a distribution concentrated on only 10000 states, which may bias the AI&M search to parameters with many zero entries.

The time complexity of the AI&M procedure shows similar characteristics as the EM procedure. For the Asia network with a sample of size 1000, the time for running AI&M was approximately 5.2 seconds, and scaled linearly in the sample size. As for EM, the crucial factor is the complexity of probabilistic inference in the Bayesian network, which AI&M must perform in order to compute the probabilities entering the *CE*-function.

We close this section by remarking that for Bayesian networks an alternative to AI&M exists in learning a Bayesian network $B'$ that extends the given complete data network $B$ with observation nodes (cf. Figure 1). The problem then is to learn the structure of the connections between primary and observation nodes, and the parameters of the model. We have also implemented this approach based on the Hugin implementation of the PC algorithm. For the smaller networks the results obtained with this method tended to be slightly inferior to AI&M. More seriously, however, the approach broke down for the Alarm network, because the junction tree for the augmented networks tended to contain large cliques, making inference, and hence EM-base parameter learning, extremely slow.

## 5 CONCLUSION AND FUTURE WORK

We have introduced the AI&M procedure for optimizing the profile(sat)-likelihood, which provides an approach to learning from incomplete data under no assumptions on the coarsening mechanism. Like EM, AI&M is a general algorithmic paradigm that can be instantiated over different types of probabilistic models. We have further proposed a particular instantiation of the AI&M procedure for learning parameters in Bayesian networks. Our results indicate that with AI&M one can obtain more accurate results than with EM when learning from non-*mar* data, especially in cases where our approximate implementation of the AI step does not introduce too large an error. That said, it must be born in mind that AI&M cannot overcome the fundamental problem that $L_{sat}$ may have many global maxima, and the true parameter $\theta$ may not be identifiable.

Ongoing work is directed at improving the quality of the AI step for Bayesian networks. In particular, one research goal is to find a method for performing the minimization in the AI step directly on the level of sufficient statistics for the subsequent M step (as usual in the E step of EM), rather than on the level of full data completions. A second topic of ongoing and future work is the use of AI&M as one core computational procedure in a likelihood ratio test for testing a "*car and* $\theta \in \Theta$" hypothesis (cf. Section 2).

## References

[1] R.G. Cowell. Parameter estimation from incomplete data for Bayesian networks. In *Proceedings of the 7th International Workshop on Artificial Intelligence and Statistics*, pages 193–196, 1999.

[2] A. P. Dawid and J. M. Dickey. Likelihood and Bayesian inference from selectively reported data. *Journal of the American Statistical Association*, 72(360):845–850, 1977.

[3] R. D. Gill, M. J van der Laan, and J. M. Robins. Coarsening at random: Characterizations, conjectures, counter-examples. In *Proceedings of the First Seattle Symposium in Biostatistics: Survival Analysis*, Lecture Notes in Statistics, pages 255–294. Springer-Verlag, 1997.

[4] D. F. Heitjan and D. B. Rubin. Ignorability and coarse data. *The Annals of Statistics*, 19(4):2244–2253, 1991.

[5] M. Jaeger. Ignorability for categorical data. *The Annals of Statistics*, 33(4):1964–1981, 2005.

[6] R. J. A. Little and D. B. Rubin. *Statistical Analysis with Missing Data*. John Wiley & Sons, 1987.

[7] R.J.A. Little. Models for nonresponse in sample surveys. *Journal of the American Statistical Association*, 77(378):237–250, 1982.

[8] C.F. Manski. *Partial Identification of Probability Distributions*. Springer, 2003.

[9] C.F. Manski. Partial identification with missing data: Concepts and findings. *International Journal of Approximate Reasoning*, 39:151–165, 2005.

[10] E.V. Nordheim. Inference from nonrandomly missing categorical data: An example from a genetic study on Turner's syndrome. *Journal of the American Statistical Association*, 79(388):772–780, 1984.

[11] M. Ramoni and P. Sebastiani. Robust learning with missing data. *Machine Learning*, 45(2):147–170, 2001.

[12] M. Ramoni and P. Sebastiani. Learning Bayesian networks from incomplete databases. In *Proceedings of UAI–97*, pages 401–408, 1997.

[13] D.B. Rubin. Inference and missing data. *Biometrika*, 63(3):581–592, 1976.

[14] C.F.J. Wu. On the convergence properties of the EM algorithm. *Annals of Statistics*, 11:95–103, 1983.

[15] M. Zaffalon. Conservative rules for predictive inference with incomplete data. In *Proceedings of the Fourth International Symposium on Imprecise Probabilities and Their Applications*, pages 406–415, 2005.